\documentclass[fleqn,usenatbib]{mnras}

\usepackage{newtxtext,newtxmath}

\usepackage[T1]{fontenc}

\DeclareRobustCommand{\VAN}[3]{#2}
\let\VANthebibliography\thebibliography
\def\thebibliography{\DeclareRobustCommand{\VAN}[3]{##3}\VANthebibliography}

\usepackage{graphicx}	
\usepackage{amsmath}	

\title[MeerKAT, e-MERLIN,\, \textit{Swift} and H.E.S.S., observations of three localised FRBs]{A MeerKAT, e-MERLIN, H.E.S.S.\ and \textit{Swift} search for persistent and transient emission associated with three localised FRBs}

\author[Chibueze et al.]{J.~O. Chibueze,$^{1,2}$\thanks{james.chibueze@nwu.ac.za}
M. Caleb,$^{3,4}$\thanks{manisha.caleb@manchester.ac.uk}
L. Spitler,$^{5}$
H. Ashkar,$^{6,17}$
F. Sch\"ussler,$^{6}$
B.~W. Stappers,$^{4}$
\newauthor
C. Venter,$^{1}$
I. Heywood,$^{7,8,9}$
A.~M.~S. Richards,$^{3}$
D.~R.~A. Williams,$^{3}$
% ...
\\ \\
}

\date{Accepted XXX. Received YYY; in original form ZZZ}

\pubyear{2021}

\begin{document}
\label{firstpage}
\pagerange{\pageref{firstpage}--\pageref{lastpage}}
\maketitle

\begin{abstract}
We report on a search for persistent radio emission from the one-off Fast Radio Burst (FRB) 20190714A, as well as from two repeating FRBs, 20190711A and 20171019A, using the MeerKAT radio telescope. For FRB 20171019A we also conducted simultaneous observations with the High Energy Stereoscopic System (H.E.S.S.) in very high energy gamma rays and searched for signals in the ultraviolet, optical, and X-ray bands. For this FRB, we obtain a UV flux upper limit of $1.39 \times 10^{-16}~{\rm erg\,cm^{-2}\,s^{-1}}$\AA$^{-1}$, X-ray limit of  $\sim6.6  \times 10^{-14}~{\rm erg\,cm^{-2}\,s^{-1}}$ and a limit on the very-high-energy gamma-ray flux $\Phi(E>120\,{\rm GeV}) < 1.7\times 10^{-12}\,\mathrm{erg\, cm^{-2}\,s^{-1}}$.
We obtain a radio upper limit of $\sim$15$\mu$Jy beam$^{-1}$ for persistent emission at the locations of both FRBs 20190711A and 20171019A, but detect diffuse radio emission with a peak brightness of $\sim$53$\mu$Jy\,beam$^{-1}$ associated with FRB 20190714A at $z=0.2365$. This represents the first detection of the radio continuum emission potentially associated with the host (galaxy) of FRB 20190714A, and is only the third known FRB to have such an association. 
Given the possible association of a faint persistent source, FRB 20190714A may potentially be a repeating FRB whose age lies between that of FRB 20121102A and FRB 20180916A. A parallel search for repeat bursts from these FRBs revealed no new detections down to a fluence of 0.08\,Jy\,ms for a 1 ms duration burst.
\end{abstract}




\section{Introduction}

Fast radio bursts (FRBs) are luminous transients that last for microseconds to milliseconds and occur at extragalactic to cosmological distances \citep[e.g.][]{LBM+07, TSB+13, MPM+20}. The estimated high radio luminosities and associated brightness temperatures required to produce these short-timescale energetic events at such distances are what makes them intriguing \citep{PHL+21, CK+21}. They have been observed to emit from $\sim 110 \, \rm{MHz} - 8 \, \rm{GHz}$, though not yet across a wide and continuous frequency band due to the variable band-limited spectra of the single pulses. Over 600 FRBs have been discovered\footnote{\url{https://www.wis-tns.org/}} of which $\sim 20$ have been seen to repeat, and it is presently uncertain whether they all do \citep{CSR+19, JOF+20}. The extraordinary observed characteristics of the repeating and non-repeating FRBs have led to various progenitor models with the bulk of them favouring neutron stars. Progenitor theories include binary neutron star mergers and collisions \citep{Totani13, YTK+18}, giant pulses from extragalactic pulsars \citep{CW+16, PP+16}, hyperflares and giant flares from magnetars \citep{PP13,PPP+18}, binary white dwarf mergers \citep{KIM13}, neutron star ``combing" \citep{Zhang18} and interactions of neutron stars with active galactic nuclei \citep{VRB+17} (see \citet{PWW+19} for a list of potential progenitors). Some of these models predict radio afterglows accompanying an FRB with timescales of days to years. \citet{LRL+16} propose that the merger of a Kerr-Newman black hole binary is one of the plausible central engines for FRBs and their afterglows. \citet{DWY17}, however, suggest that the persistent emission is due to an ultra-relativistic pulsar wind nebula sweeping up its
ambient medium with FRBs repeatedly produced through one of several potential mechanisms. In the magnetar model by \citet{MBM19}, FRBs produced by binary neutron star mergers and accretion induced collapse are expected to be accompanied by persistent radio continuum emission on timescales of months to years. The persistent emission is powered by the nebula of relativistic electrons and magnetic fields inflated by the magnetar flares \citep{MBM19}. The existence of persistent emission associated with FRBs could provide vital clues to their origin. Moreover,  potential candidates and models for FRB progenitors predict counterparts in the X-ray an TeV bands. For example, a model by~\citet{10.1093/mnrasl/slu046} predicts millisecond outbursts of TeV emission accompanying FRBs from magnetars. In 2020, FRB\,20200428 was discovered for the first time from a galactic magnetar, SGR\,1935+2154. Furthermore, an X-ray counterpart to this FRB was deteced for the first time by several instruments~\citep{tavani2020xray,Ridnaia:2021,INTEGRAL_BURST_A,HXMT_Bursts}.

Of the 19 FRBs that have been associated with host galaxies\footnote{\url{https://frbhosts.org/}}, only the sub-arcsecond localisation of the repeating FRB 20121102A to a host galaxy at a redshift of $z = 0.19273 \pm 0.0008$ \citep{SBC+17, BTA+17} showed that it is physically associated with a compact ($\leq$ 0.7 pc), persistent radio source of luminosity $\nu L_{\nu} \sim 10^{39}$ erg s$^{-1}$ at a few
GHz \citep{MPH+17}. This source is detectable from 300\,MHz -- 26\,GHz \citep{RVI20, CLW+17} and is seen to exhibit $\sim 10\%$ variability on day timescales. In contrast, a similar sub-milliarcsecond localisation of another repeating FRB\,20180916B to a nearby massive spiral galaxy at $z = 0.0337 \pm 0.0002$ \citep{MNH+21} showed no associated persistent radio emission. This places a strong upper limit on the persistent source luminosity of $\nu L_{\nu} \lesssim 7.6 \times 10^{35}$ erg s$^{-1}$ at 1.6\,GHz, which is three orders of magnitude lower than that of FRB 20121102A. 
Recently, the CHIME/FRB collaboration announced heightened activity in the repeating FRB 20201124A \citep{CHIME/FRB+21}, which was localised to a host galaxy at a redshift of $z=0.0979 \pm 0.0001$ \citep{FDL+21}. Persistent radio emission was detected by the upgraded Giant Metrewave Radio Telescope (uGMRT) \citep{WBG+21} and the Karl G. Jansky Very Large Array (JVLA) \citep{RPP+21} on angular scales of a few arcseconds, but resolved out to scales of $\sim0.1$~arcseconds with the European VLBI Network \citep{MKH+21}. 

Localisations of four one-off FRBs through imaging of buffered raw voltage data at 1.4\,GHz \citep{BDP+19, PMM+20, MPM+20} by the Australian SKA Pathfinder (ASKAP) telescope did not yield persistent radio continuum emission from the host galaxies \citep{BSP+20}. 
Australian Telescope Compact Array (ATCA) observations of FRBs 20180924B, 20181112A, 20190102C and 20190608B were conducted at a centre frequency of 6.5\,GHz. No persistent emission as luminous as the one associated with FRB 20121102A was detected for the ASKAP FRBs \citep{BSP+20}. While the true age of FRB~121102A is unknown, models based on polarization studies predict the age to be $\sim 6 - 17$ years \citep{HMS+21}. It is possible that younger, more active FRBs like FRB 20121102A are associated with persistent radio emission while the emission might have faded over time for the older ones. The possibility of repeating FRBs not being so uncommon after all \citep{Ravi19} 
along with the increasing arcsecond localisations suggests that we are entering an era where we can begin to look for evidence of multiple classes by studying FRB host galaxies and multi-wavelength counterparts. 

In this paper, we report on the search for persistent radio emission in the host galaxies of one apparent one-off source (FRB 20190714A) and two repeating sources (FRBs 20171019A and 20190711A) \citep{KSO+19, KSF+21} using MeerKAT. In case of the latter, we also conducted simultaneous observations with the High Energy Stereoscopic System (H.E.S.S.) in very high energy gamma rays. In addition, we searched for signals in the ultraviolet, optical, and X-ray bands. The paper is structured as follows. In Section~\ref{sec:data}, we discuss our observations and data reduction; in Section~\ref{sec:results}, we discuss the single radio continuum detection and derived multi-wavelength upper limits. Our discussion and conclusions follow in Section~\ref{sec:discussion} and~\ref{sec:conclusions}. 


\section{Observations and data reduction}\label{sec:data}

\subsection{MeerKAT observations}

The MeerKAT 64-parabolic-dish array \citep{2016mks..confE...1J,2020ApJ...888...61M} is located in the Northern Karoo desert near Carnarvon, South Africa. Each  ``offset Gregorian" parabolic dish antenna has an effective diameter of 13.5\,m. The inner core of the array contains 48 of the 64 dishes in a 1\,km radius, while the remaining 16 dishes are spread outward up to 8\,km. The shortest and longest baselines of the MeerKAT array are 29\,m and 8\,km, respectively, providing angular scales of 5$''$ to 27$'$ at the central frequency, of the L-band receiver used here, of 1283\,MHz. Multi-epoch observations of the FRB fields were conducted with the MeerKAT array (Project ID: SCI-20190418-VC-01) at L-band (856\,MHz to 1712\,MHz). Details of the MeerKAT observations are presented in Table \ref{tab:1}. Only Stokes I (total intensity) of the MeerKAT observations are considered in this paper.
The data correlation was done with the SKARAB correlator~\citep{2016JAI.....541001H} in 4k mode which gives 4096 channels across the 856\,MHz bandwidth resulting in a frequency resolution of $\sim$209 kHz. The data were reduced using the semi-automated MeerKAT data analysis pipelines - $oxkat$\footnote{\url{https://ascl.net/code/v/2627}} \citep{oxkat20}.

\subsubsection{Imaging analysis}

The $oxkat$ pipeline employs a collection of publicly available radio interferometry data reduction software. The final data products, including reduced and calibrated visibility data (including self-calibration), continuum (including sub-band) images as well as diagnostic plots, are provided by the pipeline. The customary configuration of the $oxkat$ pipeline incorporates flagging, cross-calibration and self-calibration processes. In the flagging process, the low-gain bandpass edges (856\,MHz to 880\,MHz and 1658\,MHz to 1800\,MHz) are flagged on all baselines, along with the location of the Galactic neutral hydrogen line at 1419.8\,MHz to 1421.3\,MHz. Several other radio frequency interference (RFI) prone regions of the spectrum are then flagged on baselines shorter than 600 m. Then, other possible RFI affected data are flagged out using the CASA routines rflag and tfcrop for the calibrators, and using the tricolour package  for the target fields.

The cross-calibration steps using $oxkat$ were standard, including setting the flux scale and deriving corrections for residual delay calibration, bandpass and time-varying gain. 
The $oxkat$ pipeline uses the customary tasks from the CASA \citep{2007ASPC..376..127M} suite for cross-calibration. After applying all the corrections to the target field, we channel-averaged the dataset by a factor of five channels before splitting out the science target. This is consistent with our science goals, since the relic sources we target are in the central parts of our fields, reducing the effect of smearing through the channel averaging. 
To deconvolve and image the target data, the WSClean imager \citep{2014MNRAS.444..606O} was used, with the multiscale and wideband deconvolution algorithms enabled to better allow imaging of the diffuse emission present in the our fields. Deconvolution was performed in ten sub-band images of each 82\,MHz wide-band. WSClean generates the multi-frequency synthesis (MFS) map, in joined-channel deconvolution mode, with a central frequency of 1283\,MHz. In other words, the MFS map is a full bandwidth map. In WSClean, each of the sub-bands is deconvolved separately with an initially high mask of 20$\sigma_{\rm rms}$ (using the auto masking function provided by WSClean), to generate an artefact-free model of the target field for the self-calibration process. This masking threshold was iteratively reduced to a value of 3$\sigma_{\rm rms}$ in the final iteration of imaging. The $oxkat$ pipeline uses the customary tasks from the Cubical software \citep{2018MNRAS.478.2399K} for self-calibration. 

\subsubsection{Single pulse searches}

In addition to obtaining correlated data, the output data stream of the F-engine are captured, delay corrected, phased and channelised before being sent over the central beamforming network to the beamforming User Supplied Equipment (FBFUSE) that was designed and developed at the Max Planck Institute for Radio Astronomy in Bonn. For this project, FBFUSE combined the data into 764 total-intensity tied-array beams which were used to populate the primary beam of $\sim 1$ deg$^{2}$ of the array. The data are then captured at 306.24 $\upmu$s time resolution by the Transient User Supplied Equipment (TUSE), a real-time transient detection backend instrument developed by the MeerTRAP\footnote{\url{https://www.meertrap.org/}} team at the University of Manchester. More details on TUSE will be presented in an upcoming paper (Stappers et al. in prep).
The GPU-based single pulse search pipeline AstroAccelerate\footnote{\url{https://github.com/AstroAccelerateOrg/astro-accelerate}} \citep{DimoudiWesley, Karel3, Karel2, Dimoudi, Karel} was used to search for bursts in real-time after incoherently de-dispersing the data in the DM range 0--5118.4~pc cm$^{-3}$ \citep[see][for more details]{CSA+20}. 

\subsection{e-MERLIN Observations}

To constrain the position of the persistent continuum emission associated with FRB\,20190714A, we conducted L-band (centre frequency of 1.51\,GHz) observations of the target with the enhanced Multi-Element Remote-Linked Interferometer Network,
e-MERLIN array in the United Kingdom (project code: CY10003) on 13 January, 2021 (see Section \ref{sec:PRS}). Six antennas were used including the 75-m Lovell telescope and the target pointing centre was R.A. $=$ 12$^h$15$^m$55$^s$.12, Dec. $= -13^{\circ}$01$\arcmin$15$\farcs$7. 1407$+$2827 was used as the bandpass calibrator, 1331$+$3030 as the
flux calibrator and 1216$-$1033 as the phase calibrator. The angular separation between the target and the phase calibrator is 2.47$^{\circ}$. The data reduction was done following standard e-MERLIN calibration procedures\footnote{\url{https://github.com/e-merlin/eMERLIN_CASA_pipeline}} with additional flagging of bad visibilities followed by imaging. We found two confusing sources in the field, at R.A. $=$ 12$^h$15$^m$44$^s$.669, Dec. $= -12^{\circ}$57$\arcmin$59$\farcs$56 and R.A. $=$ 12$^h$15$^m$37$^s$.216, Dec. $= -13^{\circ}$09$\arcmin$33$\farcs$44 at 4.1\arcmin~and 9.4\arcmin~from the pointing centre, respectively. They had apparent flux densities of 4 and 1.3\,mJy without primary beam correction. We used these for self-calibration of the field and then subtracted them before final imaging. The final image synthesized beam is $0\farcs65\times0\farcs15$, position angle $15^{\circ}$ elongated in the Declination direction due to the low target elevation from the UK.


\subsection{The \textit{Swift} satellite: UVOT and XRT observations}
Neil Gehrels Swift Observatory (\textit{Swift}) is a multi-wavelength NASA space mission operating in soft-X-rays and optical/UV.  
Here we use data from the X-ray Telescope (XRT)~\citep{Swift_XRT} which operates in the soft X-ray domain of $0.3 - 10$ keV as well as data taken by the UV/Optical Telescope (UVOT)~\citep{Swift_UVOT} operating in the UV to optical domain ($170 - 600$ nm). During the FRB 20171019A multi-wavelength (MWL) observing campaign, two 2~ks target-of-opportunity (ToO) observations were performed with \textit{Swift} from 2019-09-28 18:37:02 to 2019-09-28 21:52:54 and 2019-10-18 18:03:00 to 2019-10-18 20:03:00 on the FRB 20171019A localisation region. 
Simultaneously with Swift-XRT, five UVOT images were taken with the UVM2 filter (central wavelengh = 2246 {\AA}) over the 2 epochs with a total exposure of 4 ks. The images are aspect-corrected and summed with the $\texttt{uvotimsum}$ tool ($\texttt{HEASOFT 6.26}$). 
Observations were performed with Swift-XRT in the standard Photon Counting observing mode (PC). The XRT PC data are processed with $\texttt{xrtpipeline}$ ($\texttt{HEASOFT 6.26}$). A summed image is extracted with $\texttt{xselect}$. 

\subsection{Very-high energy gamma-ray observations with H.E.S.S. }
Observations of FRB 20171019A were also obtained in the very-high energy gamma-ray domain with the H.E.S.S.\ imaging atmospheric Cherenkov telescope array, sensitive in the range between a few tens of GeVs and 100 TeV. H.E.S.S.\ is located on the Khomas Highland plateau of Namibia (23$^{\circ}16'18''$ South, $16^{\circ}30'00''$ East), at an elevation of $\sim$1800 m above sea level. Observations took place contemporaneously to the first epoch of MeerKAT observations of FRB 20171019A described above. The data set was obtained with the H.E.S.S.\ phase II array, including the upgraded 12\,m-diameter CT1-4 telescopes~\citep{HESS1U:2020} and the large 28\,m-diameter CT5 telescope~\citep{HESS2_CAMERA2014}. A standard data quality selection was applied to the data~\citep{Aharonian2006a}. The events have then been selected and their direction and energy reconstructed using a log-likelihood minimization comparing the recorded shower images of all triggered telescopes (requiring at least two telescopes to see the same gamma-ray event) to a semi-analytical model of air showers~\citep{de-Naurois2009a}. 

We define a circular region-of-interest centered on the position of FRB 20171019A with a radius of 0.12$^{\circ}$, optimal for a point-like source of emission as expected from FRB\,20171019A. The background level in this ON region was determined using the standard ``ring background'' technique~\citep{RingBg} based on a radially symmetric ring around the source position. This technique allows us to derive the background level from the same field of view and assures that the gamma-ray signal and background are estimated with the same acceptance and under the same observation conditions. 

\section{Results}\label{sec:results}

\subsection{MeerKAT}

The theoretical thermal noise of the MeerKAT can be calculated as
\begin{equation}
S_{\rm{rms}} = \frac{1}{\eta_c} \frac{\rm{SEFD}}{\sqrt{n_{\rm{pol}}\times N(N-1)\times\Delta\nu\times t_{\rm{int}}}}.
\end{equation}
The system equivalent flux density (SEFD) of MeerKAT at the 1.28\,GHz is 443\,Jy and $\eta_c$ is the correlator efficiency. We used  $n_\mathrm{pol}$ = 2 polarisation products (XX and YY), N = 64 telescopes, $\Delta\nu$ = 856\,MHz bandwidth and $t_\mathrm{int}$ = 21600 sec observing time for one epoch. This gives the theoretical rms of $\sim$ 2~$\mu$Jy beam$^{-1}$. 
The typical image rms obtained from our residual images is $\sim$\,5\,$\mu$Jy beam$^{-1}$, which is 2.5 times the expected theoretical rms.
The wideband MFS image does not allow primary beam correction procedure as this can only be done on the sub-band images with limited rms for detection of the sources. However, our sources are the phase centres of our fields and thus unaffected by the effect of the primary beam.

Due to the lack of MeerKAT primary beam correction, we did not compare the flux densities of the discrete sources with their NRAO (National Radio Astronomy Observatory) VLA (Very Large Array) Sky Survey (NVSS) counterparts. However, Chibueze et al.\ (2021, submitted) confirmed that the overall flux densities obtained with MeerKAT and NVSS are in good agreement with each other within errors of $\sim 5$\%. We compared the astrometry of the discrete radio sources obtained with MeerKAT and NVSS in Figure~\ref{fig:nvss_meerkat_astrometry}. The position uncertainty of the MeerKAT ranges from 0$\farcs$2 (close to the centre of the primary beam) to a few arcseconds towards the edge of the primary beam. The scatter observed in Figure~\ref{fig:nvss_meerkat_astrometry} is mostly due to the probability of the centroids of emission in the $\sim$45\arcsec~NVSS resolution being different from the centroids at MeerKAT's resolution and partly due to higher position uncertainty of the fainter sources. Therefore, we conclude that our MeerKAT data are well calibrated and the flux density and astrometry are as accurate as the errors indicate.

\subsubsection{Looking for persistent continuum emission associated with the FRB fields}

\begin{figure*}
\begin{center}  
\includegraphics[width=0.8\textwidth]{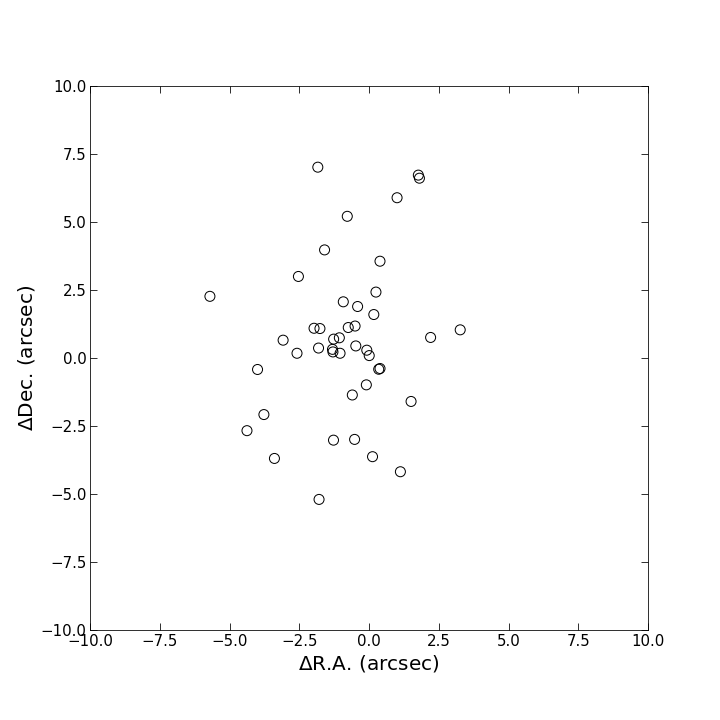}
\end{center}
\caption{
Astrometric comparison between MeerKAT and NVSS discrete compact sources.
 The open circles represent the difference in position between the MeerKAT and NVSS sources.
\label{fig:nvss_meerkat_astrometry}
}
\end{figure*}
\begin{figure*}
\begin{center}
\includegraphics[width=0.77\textwidth]{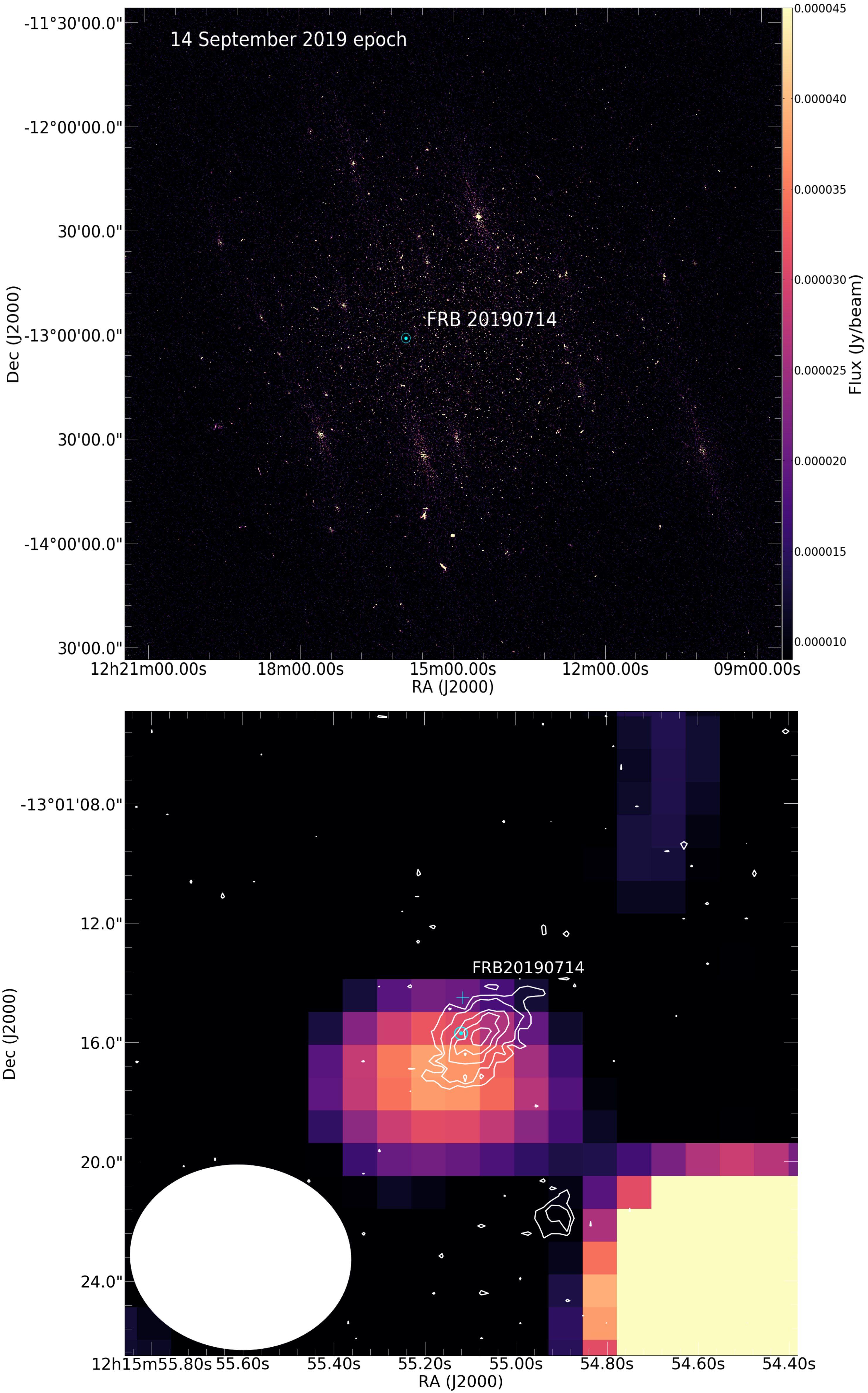}
\end{center}
\caption{
FRB 20190714A MeerKAT epoch I image (top) and a zoom-in (bottom) around the position of the FRB indicated by the cyan circle. White contours (levels: 300, 500, 900, 1200, 1600 counts) represent the PanSTARRS $i$-band optical counterpart coincident in position with the persistent radio emission. The white ellipse in the bottom left corner represents the beam size of MeerKAT. The cyan cross indicates the position of the detected compact emission in our e-MERLIN observations. 
\label{fig:190714_e1}
}
\end{figure*}
\begin{figure*}
\begin{center}
\includegraphics[width=0.8\textwidth]{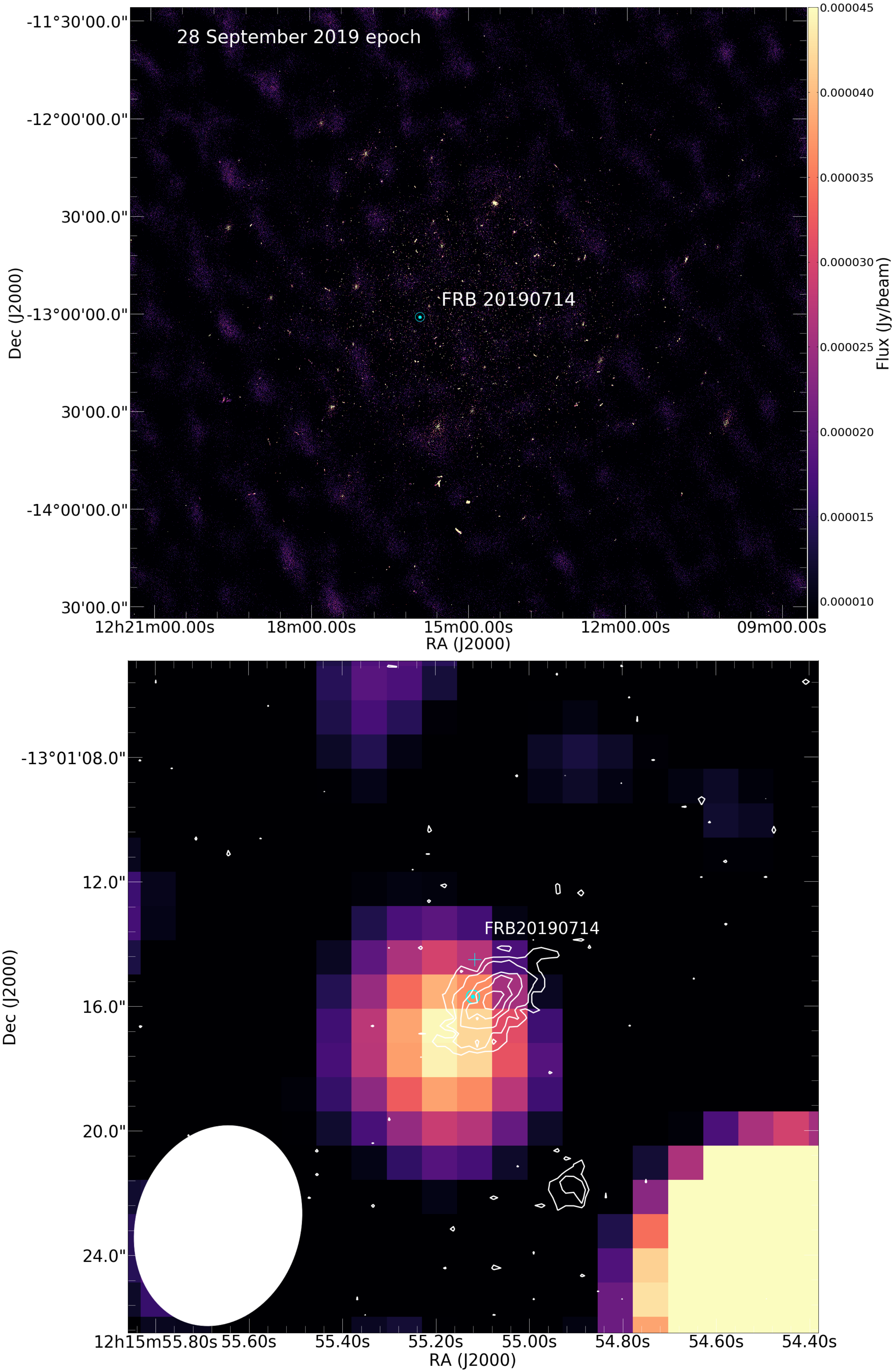}

\end{center}
\caption{
FRB 20190714A MeerKAT epoch II image (top) and a zoom-in (bottom) around the position of the FRB indicated by the cyan circle. White contours (levels: 300, 500, 900, 1200, 1600 counts) represent the PanSTARRS $i$-band optical counterpart coincident in position with the persistent radio emission. The white ellipse in the bottom left corner represents the beam size of MeerKAT. The cyan cross indicates the position of the detected compact emission in our e-MERLIN observations. 
\label{fig:190714_e2}
}
\end{figure*}

Considering the results of the astrometric comparison with NVSS (see Figure~\ref{fig:nvss_meerkat_astrometry}), we considered potential associations of continuum sources in the MeerKAT observations with the FRB location to sources within 5$^{\prime\prime}$. Using this spatial coincidence criterion, we identified a persistent 1283\,MHz continuum source near FRB 20190714A, detected in both the 14 September 2019 and the 28 September 2019 epoch. The peak of the MeerKAT radio emission is offset by $\sim$\,2$^{\prime\prime}$.1 from the peak of the $i$-band magnitude of the optical galaxy identified in the Panoramic Survey Telescope and Rapid Response System (PanSTARRS, located at Haleakala Observatory) image (shown as contours in Figures \ref{fig:190714_e1} and \ref{fig:190714_e2}). The MeerKAT radio source is offset by 1$\farcs$68 from the localisation region of FRB 20190714 (cyan circle in Figures \ref{fig:190714_e1} and \ref{fig:190714_e2}).

\subsubsection{e-MERLIN detection of compact emission towards FRB\,20190714}
\label{sec:PRS}

Compact persistent emission was detected in the 1.51\,GHz
e-MERLIN image at R.A. $=$ 12$^h$15$^m$55$^s$.116, Dec. $= -13^{\circ}$01$\arcmin$14$\farcs$48 at 86\,$\mu$Jy\,beam$^{-1}$ by e-MERLIN. The stochastic position uncertainty is (0.04, 0.15) arcsec and the  uncertainty (due to the separation between phase-calibrator and target, and antenna position uncertainty) is (0.013, 0.056) arcsec, giving a total astrometric uncertainty of (0.04, 0.16) arcsec in R.A. and Dec., respectively. The offset from the FRB position is negligible in R.A. and 1.2 arcsec in Dec.
The rms in this region (of full primary beam sensitivity) is 20\,$\mu$Jy\,beam$^{-1}$, making this a 4.3$\sigma_{\mathrm{rms}}$ detection.  It is $\sim$1.5$\sigma_{\mathrm{rms}}$ higher than that of the MeerKAT detection. Although the e-MERLIN flux scale nominal uncertainty is $\sim$5\%, in these data it is possibly higher due to the low declination of the phase-reference source and to the strong RFI which were removed from the data but may have affected the linearity of the receiver response. The peak of the e-MERLIN radio emission is offset by $\sim1\farcs4$ from the peak of the PanSTARRS $i$-band emission in Figures \ref{fig:190714_e1} and \ref{fig:190714_e2}. The e-MERLIN radio source (shown by the cyan cross in Figures \ref{fig:190714_e1} and \ref{fig:190714_e2}) is offset by 0$\farcs$53 from the localised position of FRB 20190714.

We estimate the probability of a chance alignment of a background persistent radio source and the host galaxy, following the procedure of \citet{EBWB18}. Instead of using the FRB localisation region, we use the area of the galaxy, which is taken as 2$^{\prime\prime}$ $\times$ 2$^{\prime\prime}$, twice the half light radius from \citet{Heintz}. Given the source has a flux density of $\sim90\mu$Jy we estimate the chance alignment probability of 0.0008, which corresponds to 3.4$\sigma$. The flux density threshold, assuming 3$\sigma$, for an unresolved radio source is $\sim$\,15\,$\mu$Jy. If instead we consider the probability of detecting any radio source above our flux density threshold of $15\mu$Jy, the probability of a chance alignment is, therefore, approximately 0.8\%, making the statistical significance of our detection 2.6$\sigma$. This represents the first detection of radio continuum emission associated with the host (galaxy) of FRB 20190714A (see Figure \ref{fig:190714_e1} and \ref{fig:190714_e2}). 
\subsubsection{MeerKAT non-detections}
No continuum emission was detected near FRBs 20171019A and 20190711A. As each of the images of these sources has an rms of $\sim$\,5\,$\mu$Jy beam$^{-1}$, the 3$\sigma$ intensity upper limit of any emission associated with FRBs 20171019A and 20190711A will be $\sim$\,15\,$\mu$Jy beam$^{-1}$ (see Table \ref{tab:1}).

Candidate pulses above a signal-to-noise (S/N) of 10 from the single pulse search with MeerTRAP were visually inspected offline. No new FRBs or repeat bursts from the known FRBs were detected above a fluence threshold of 0.08\,Jy\,ms assuming a 1~ms duration burst.

\subsection{\textit{Swift}}
\begin{figure*}
\begin{center}
\includegraphics[width=0.8\textwidth]{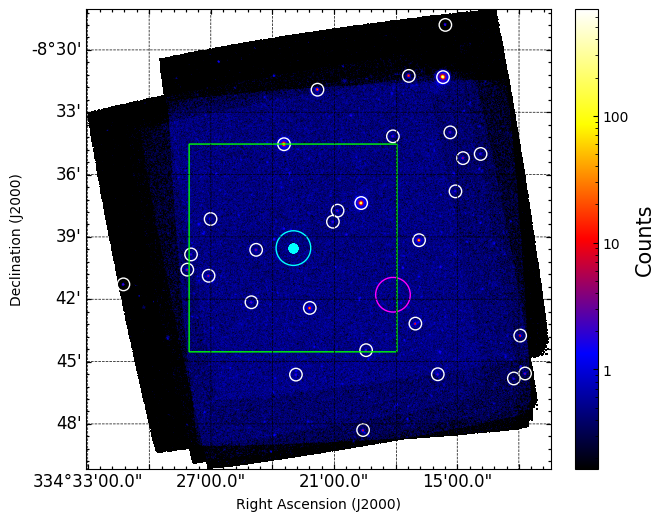}
\caption{UVOT summed image of FRB 20171019A region taken during the MWL observation campaign in September-October 2019. The white circles indicate sources detected above $5 {\sigma}$. The cyan dot denotes the location of FRB 20171019A, the circle around it indicates the region used to derive the upper limits while the magenta region indicates the background region used. The green box indicates FRB\,20171019A 90\% localisation region as reported in ~\citet{KSO+19}.}
\label{fig:UVOT_FRB171019}
\end{center}
\end{figure*}

The UVOT summed image is presented in Figure~\ref{fig:UVOT_FRB171019}. The UVOT field of view corresponds roughly to the uncertainty\footnote{\url{https://www.wis-tns.org/object/20171019a}} of the localisation region of FRB 20171019A (RA = 7.5\arcmin and DEC = 7\arcmin). Using $\texttt{uvotdetect}$, we find 30 sources above the $5 {\sigma}$ level and within the FRB 20171019A uncertainty region. Using a 3 arcsec maximum separation, which is slightly larger than the UVOT PSF~\citep{uvot_psf}, these sources are cross-matched with known catalogue sources. We find that out of the 30 sources detected by UVOT, 28 are spatially coincident with stars catalogued in the SDSS catalogue \citep[DR12;][]{SDSS-DR12}, and one source is coincident with a galaxy (AGN broadline SDSS ID: 1237652599570890948 at $z\sim0.156$). This galaxy is also detected by the MeerKAT radio observations. 
We use the NASA/IPAC Extragalactic Database (NED)\footnote{\url{https://ned.ipac.caltech.edu}; NED is funded by the National Aeronautics and Space Administration and operated by the California Institute of Technology} to search for known galaxies in the FRB 20171019A uncertainty regions. We find multiple galaxies with unknown redshifts, therefore we cannot draw conclusions on the host galaxy from our observations. Using a 50\arcsec\, circular ON region centred on the position of FRB 20171019A and a 50\arcsec\, OFF region that does not contain any of the detected sources, we run the $\texttt{uvotsource}$ tool with a $5 {\sigma}$ background threshold and obtain a flux upper limit of $1.4 \times 10^{-16}~{\rm erg\,cm^{-2}\,s^{-1}}$\AA$^{-1}$ without applying a Calactic extinction correction. 

\begin{figure*}
\begin{center}
\includegraphics[width=0.8\textwidth]{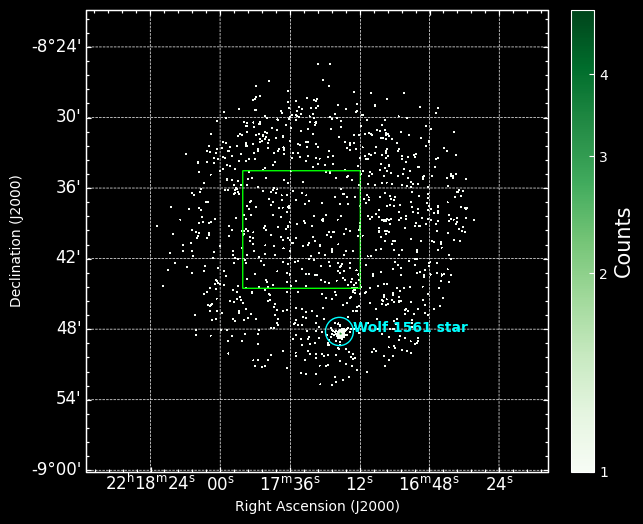}
\caption{XRT summed image of FRB 20171019A region taken during the MWL observation campaign in September - October 2019. The position of the Wolf 1561 star is shown in cyan and is labelled. The green box indicates FRB\,20171019A 90\% localisation region as reported in \citet{KSO+19}.}
\label{fig:XRT_FRB171019}
\end{center}
\end{figure*}

The XRT summed image is shown in Figure~\ref{fig:XRT_FRB171019}. At the edge of the field-of-view, we detect a source spatially coincident with the Wolf 1561 star. As we consider this source unrelated to the FRB, we use the online Swift-XRT data products generator~\citep{XRT_CURVE_1}~\citep{XRT_CURVE_2} to derive upper limits in the 0.3-10 keV range on the count rate 
of $0.001885~{\rm counts.s^{-1}}$. Using $\texttt{WebPIMMS}$\footnote{\url{https://heasarc.gsfc.nasa.gov/cgi-bin/Tools/w3pimms/w3pimms.pl}} (v4.11a) and assuming a weighted average $N_{\rm H} = 5.12\times 10^{20}~\rm{cm^{-2}}$ from the direction of the source estimated from the NASA's HEASARC~\footnote{\url{https://heasarc.gsfc.nasa.gov/cgi-bin/Tools/w3nh/w3nh.pl}} online tools~\citep{2016A&A...594A.116H} and a power law model with a photon index = 2, this upper limit translates to an energy flux of  $6.6 \times 10^{-14}~{\rm erg\,cm^{-2}\,s^{-1}}$  ($8.3 \times 10^{-14}~{\rm erg\,cm^{-2}\,s^{-1}}$ unabsorbed).

\subsection{H.E.S.S.}
\begin{figure*}
\begin{center}
\includegraphics[width=0.8\textwidth]{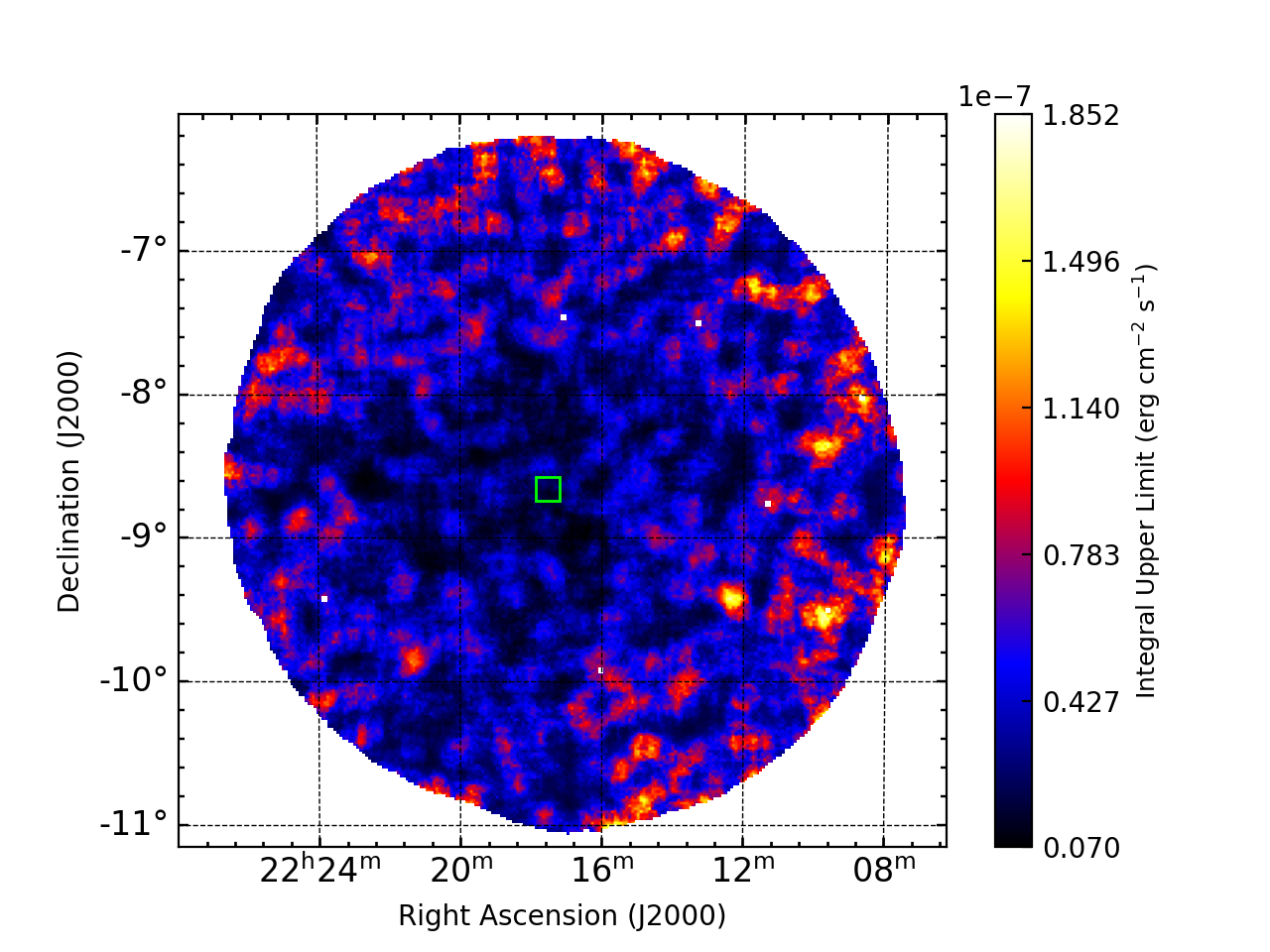}
\caption{Map of upper limits on the VHE gamma-ray energy flux derived from the H.E.S.S. observations. The limits are valid above $120\,\mathrm{GeV}$ and assume a photon flux distribution following an $E^{-2}$ dependence. The green box indicates the FRB\,20171019A 90\% localisation region as reported in \citet{KSO+19}.The oversampling radius is 0.1$^{\circ}$.}
\label{fig:HESS_ULmap}
\end{center}
\end{figure*}

No significant gamma-ray excess above the expected background is detected from the direction of FRB 20171019A, with 52 gamma candidate events from the source region and 524 background event.  A second analysis using an independent event calibration and reconstruction~\citep{Parsons2014a} confirms this result. A search for variable emission on timescales ranging from milliseconds to several minutes with tools provided in~\citep{Brun_transient_tools} does not reveal any variability above 2.2 $\sigma$. For the total data set of 1.8\,h, 95\% confidence level (C.~L.) upper limits on the photon flux are derived using the method described by \citet{Rolke2005a}.
The energy threshold of the data is highly dependent on the zenith angle of the observations. For these observations, the zenith angles range from 15 to 25 deg, which leads to an energy threshold for the stacked data set of $E_{\mathrm{th}} = 120$\,GeV. The upper limit on the Very High Energy (VHE) gamma-ray flux above that threshold and assuming an energy dependence following $E^{-2}$ is $\Phi(E>120\,{\rm GeV}) < 2.10\times 10^{-12}\,\mathrm{cm
^{-2}\,s^{-1}}$ or $\Phi(E>120\,{\rm GeV}) < 1.7\times 10^{-12}\,\mathrm{erg\,
cm^{-2}\,s^{-1}}$. A variation of $\pm$~0.5 of the assumed spectral index leads to a variation in the upper limit of less than $\pm$~19\%.  A map of energy flux upper limits covering the full region accessible within the H.E.S.S.\ field of view above 120 GeV is given in Figure~\ref{fig:HESS_ULmap}.


\begin{table*}
\caption{Details of the FRB fields observed with MeerKAT.}
\label{tab:1}
\begin{tabular}{lcccc}
\hline
Field name & Observation date & Synthesized beam & rms ($\mu$Jy\,beam$^{-1}$)& Detected? \\
\hline
FRB 20171019A & 28 September 2019 & &-- & No (calibration failure)\\
FRB 20171019A & 18 October 2019 & 6$\farcs$8 $\times$ 5$\farcs$0 & 5.2 & $<15 \mu$Jy~beam$^{-1}$\\
\hline
FRB 20190711A & 23 August 2019 & 11$\farcs$7 $\times$ 4$\farcs$9 & 4.9 & $<15 \mu$Jy~beam$^{-1}$\\
FRB 20190711A & 09 September 2019 & 12$\farcs$5 $\times$ 4$\farcs$9 & 4.6 & $<15 \mu$Jy~beam$^{-1}$\\
 \hline
FRB 20190714A & 14 September 2019 & 7$\farcs$1 $\times$ 6$\farcs$2 & 4.2 & 54.4~$\mu$Jy~beam$^{-1}$\\
FRB 20190714A & 28 September 2019 & 6$\farcs$5 $\times$ 5$\farcs$1 & 5.8 & 52.0~$\mu$Jy~beam$^{-1}$\\
 \hline 
\end{tabular}
\end{table*}

\begin{table*}
\caption{Details of the radio continuum source associated with FRB 20190714A.}
\label{tab:2}
\begin{tabular}{lcccccccc}
\hline
Field name & Observation date & Telescope & $\nu_\mathrm{centre}$ (GHz)& $\alpha_\mathrm{J2000}$ & $\delta_\mathrm{J2000}$ & Maj. $\times$ min. axis & Pos.\ angle & Int. flux density \\ \\ 
\hline
FRB 20190714A & 28 September 2019 & MeerKAT & 1.283 & 12$^h$15$^m$55$^s$.154 & -13$^\circ$01\arcmin17\farcs30 & 9\farcs6 $\times$ 7\farcs4 & 88.7$^\circ$ & 87.4 $\mu$Jy\\
FRB 20190714A & 18 October 2019 & MeerKAT & 1.283 & 12$^h$15$^m$55$^s$.193 & $-13^\circ$01\arcmin17\farcs18 & 8\farcs2 $\times$ 6\farcs4 & 12.2$^\circ$ & 80.7 $\mu$Jy \\
FRB 20190714A & 13 January 2021 & e-MERLIN & 1.510 & 12$^h$15$^m$55$^s$.116 & $-13^\circ01$\arcmin14\farcs51 & 0\farcs15 $\times$ 0\farcs65 & 17.6$^{\circ}$ & 107.5 $\mu$Jy  \\
\hline 
\end{tabular}
\end{table*}

\begin{figure*}
\begin{center}
\includegraphics[width=0.8\textwidth]{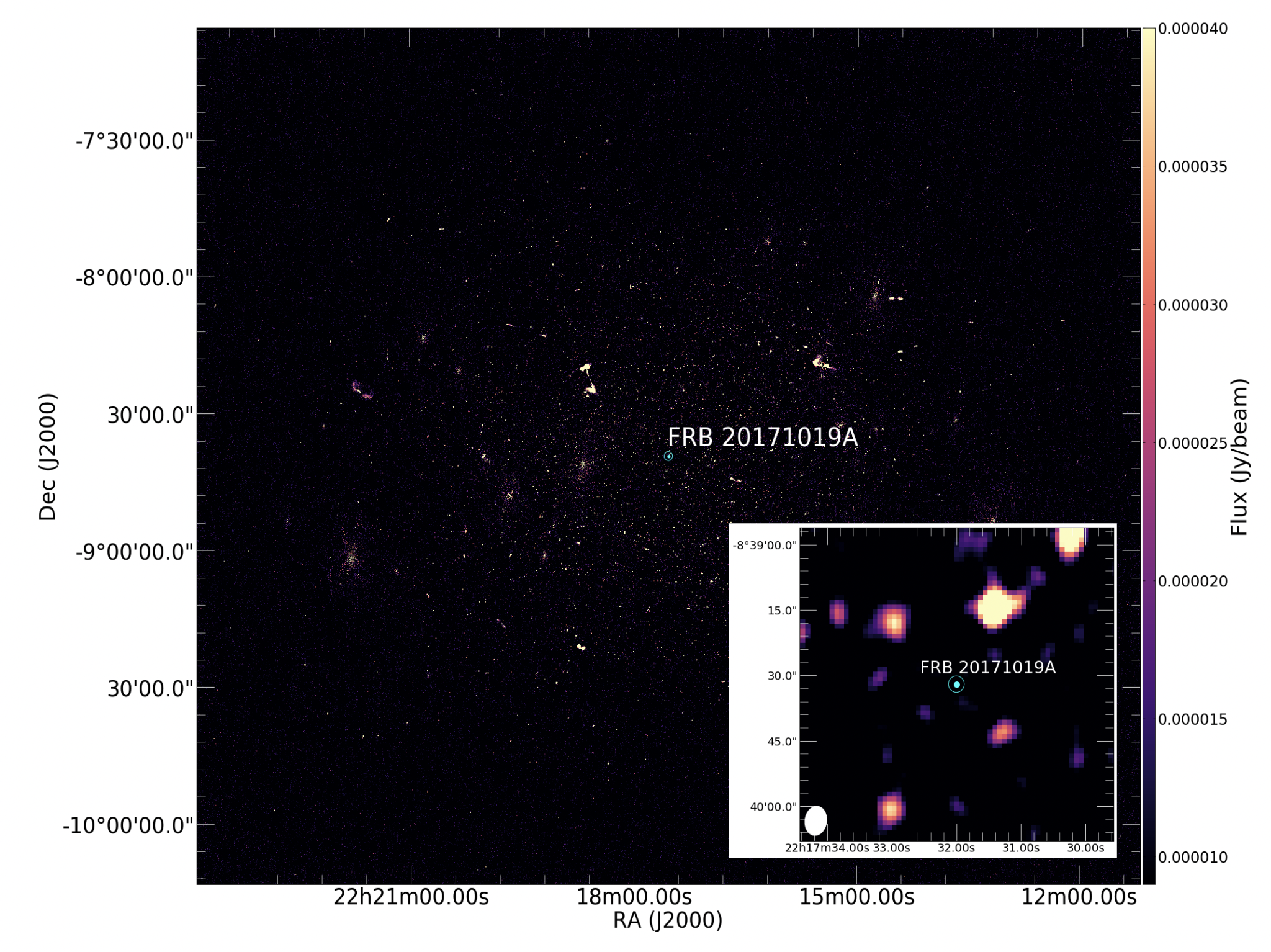}
\end{center}
\caption{
FRB 20171019A MeerKAT image and a zoom-in (insert) around the position of the FRB. The white ellipse on the bottom left corner of the insert represent the beam size of MeerKAT.  
\label{fig:171019}
}
\end{figure*}

\begin{figure*}
\begin{center}
\includegraphics[width=0.8\textwidth]{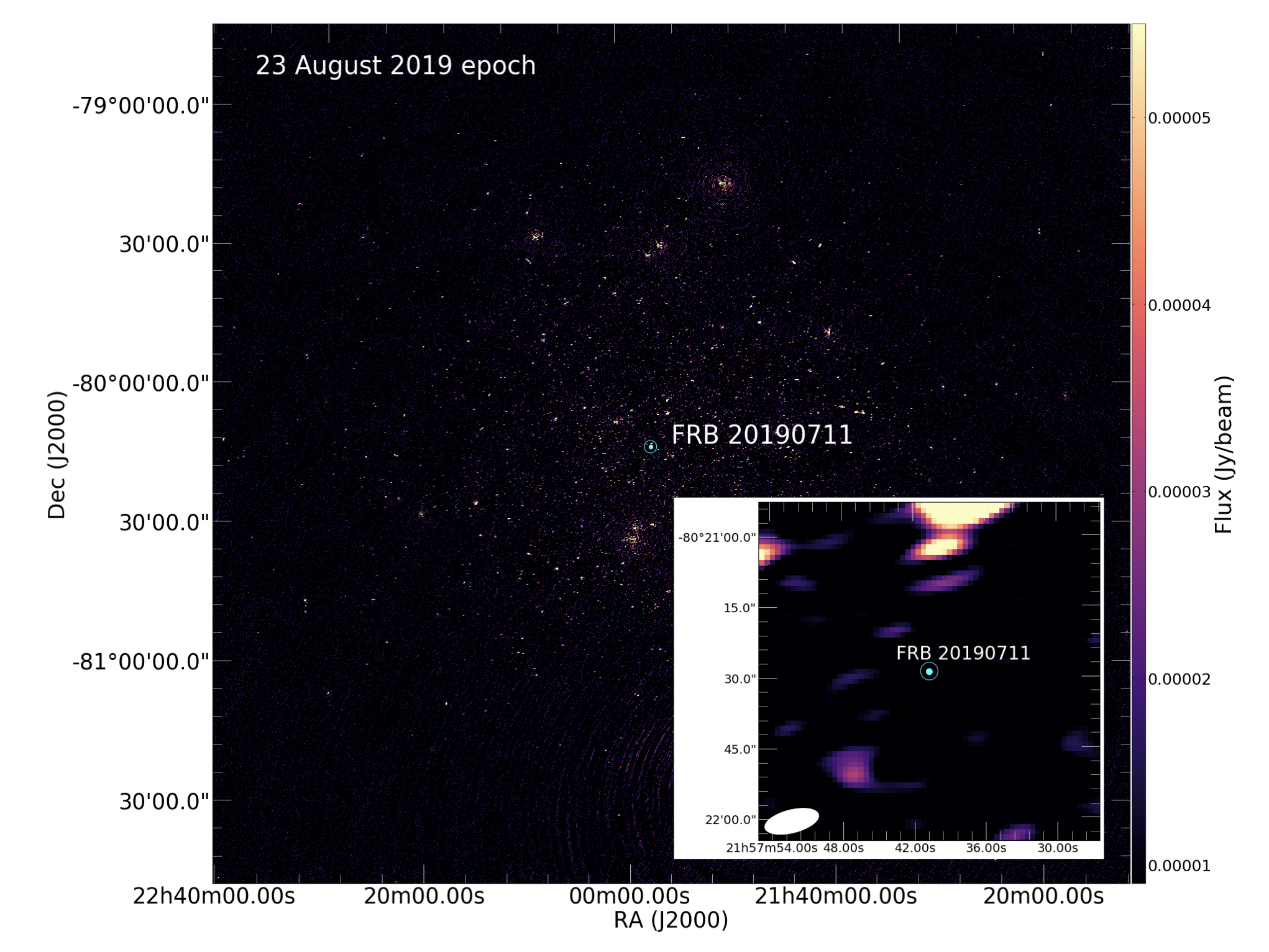}
\end{center}
\caption{
FRB 20190711A MeerKAT epoch I image and a zoom-in (insert) around the position of the FRB. The white ellipse on the bottom left corner of the insert represent the beam size of MeerKAT.  
\label{fig:190711}
}
\end{figure*}

\begin{figure*}
\begin{center}
\includegraphics[width=0.8\textwidth]{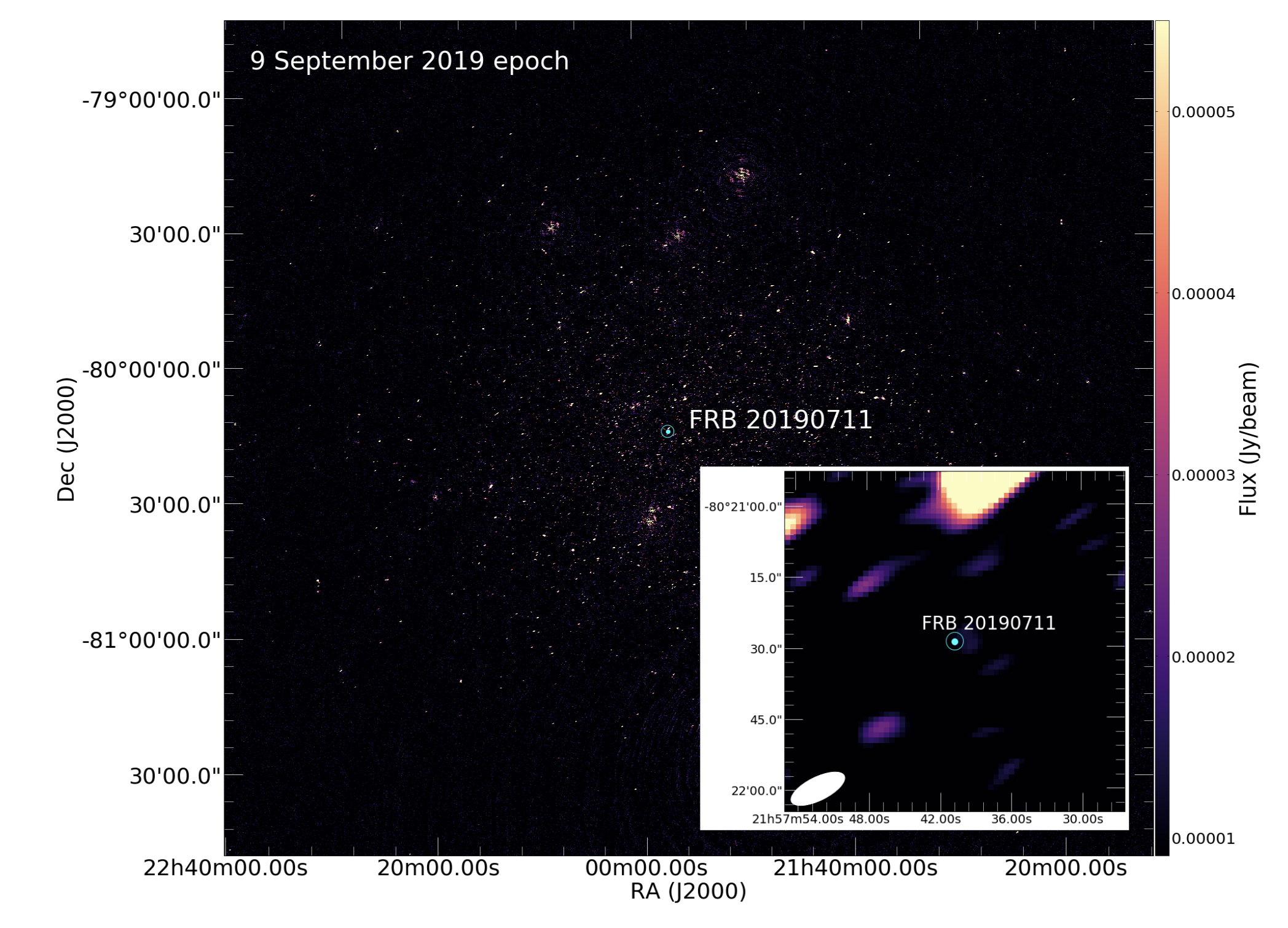}
\end{center}
\caption{
FRB 20190711A MeerKAT epoch II image and a zoom-in (insert) around the position of the FRB. The white ellipse on the bottom left corner of the insert represent the beam size of MeerKAT.  
\label{fig:190711_e2}
}
\end{figure*}

\section{Discussion}
\label{sec:discussion}


Of the targeted FRB fields reported here, only FRB 20190714A is observed to be spatially coincident with a persistent radio continuum source. We obtain an upper limit of $\sim 15\,\mu$Jy beam$^{-1}$ for FRBs 20190711A and 20171019A, respectively, and a peak intensity of $\sim53\,\mu$Jy beam$^{-1}$ for the emission coincident with FRB 20190714A. This source is detected at both epochs with similar intensities within the measured rms of the images (see Tables~\ref{tab:1} and \ref{tab:2} for details). The values in the Table \ref{tab:2} are derived by carrying out 2D Gaussian fit using similar ellipses enclosing the detected persistent emission. The average flux density is $\sim 3$ times less than that of the persistent source associated with FRBs 20121102A, one of the most prolific repeaters, located at $z = 0.19273(8)$. Persistent radio emission from FRB 20201124A was detected by the uGMRT \citep{WBG+21} and the JVLA \citep{RPP+21} on angular scales of a few arcseconds. However, it is resolved out at scales of $\sim0.1$~arcseconds with the European VLBI Network \citep{MKH+21} suggesting that it is not a compact source directly associated with the FRB. In contrast, the other localised, prolific repeating FRB 20180916A has no persistent radio counterpart. 

In the image in Figure \ref{fig:190714_e2} one can see that the persistent radio source lies at the edge of the optical extent of the host galaxy as seen in PanSTARRS observations \citep{Heintz}.
Our derived 1283\,MHz peak position with MeerKAT places it just 1\farcs68 away from the position of FRB 20190714A \citep[$\alpha_{J2000},\delta_{J2000}$ = 12$^h$15$^m$55$^s$.12, -13$^\circ$01\arcmin15\farcs70;][]{Heintz}. The positional uncertainty on the FRB position is 0\farcs283. Similarly, the peak 1.51\,GHz e-MERLIN position of the persistent radio source is separated from the position of FRB 20190714A by 0$\farcs$53. The persistent source near FRB 20190714A has a flux broadly consistent with the MeerKAT flux and is unresolved on the e-MERLIN baselines. The large offset from the centre of the galaxy makes the persistent source unlikely to be an AGN. So far this FRB has not been seen to repeat. Higher resolution imaging will be required to be certain of a direct association of the persistent source with the FRB.
We did not have sufficient sensitivity in the sub-band images, thus, we are unable to derive the spectral index of the emission of the host galaxy.

Our e-MERLIN observations probe a different spatial scale than the size of the persistent radio source associated with FRB~20121102A. At the angular diameter distance of FRB~20190714A (780 Mpc), an unresolved source with an angular size of 0$\farcs$6 corresponds to a physical extent of $\lesssim$2.3~kpc. 
The uGMRT reported the detection of an unresolved radio emission at 650\,MHz with a flux density of $700 \pm 100\, \mu$Jy \citep{WBG+21}, while the JVLA detected persistent emission with a flux density of $340 \pm 30\, \mu$Jy at 3\,GHz \citep{RPP+21}. Assuming the estimated spectral index between these frequencies \citep[$\sim-0.5$,][]{RPP+21}, the 1.3\,GHz flux density would be $\sim500\, \mu$Jy (similar to the 3-$\sigma$ upper limit on observations from $1 - 2$\,GHz; \citealt{LTC+21}). The flux density we measured for FRB 20190714A is a factor of $\sim$10 lower than FRB20201124A, but FRB 20190714A is also a factor 2.6 more distant. Therefore, the flux densities would be comparable if they were at similar distances. 

Given the resolution of MeerKAT we are unable to definitively state whether the persistent emission is associated with a star-forming region or the FRB itself. However, the increased resolution with the e-MERLIN baselines would tend to favour a compact source similar to the one observed in FRB 20121102A. One of the leading models to explain the bursts from, and radio counterpart to FRB 20121102A, is a young nebula powered flaring magnetar embedded in a 20–50 year-old supernova remnant \citep{Beloborodov, MMS19}. The lack of a bright persistent radio source associated with the repeater FRB 20180916A suggests that it is comparatively older at $\gtrsim 200-500$ years and the persistent radio source may have faded. In the model by \citet{MMS19}, the nebula is suggested to contribute significantly to the rotation measure and dispersion measure (DM), as well as to the persistent radio luminosity. These values are expected to decrease on a timescale of a few decades to centuries. Given the association of a comparatively fainter persistent source, FRB 20190714A may potentially be a repeating FRB whose age lies between that of FRB 20121102A and FRB 20180916A.
Millisecond magnetars formed through standard astrophysical channels such as hydrogen poor superluminous supernovae and long duration gamma-ray bursts are consistent with the progenitors of FRBs expected in low-metallicity dwarf galaxies with high specific star-formation rate such as for FRB 20121102A. However, \citet{MBM19} note that it is also possible to form such sources through a variety of channels, including binary neutron star mergers and accretion induced collapse of white dwarfs in environments and host galaxy demographics different to FRB 20121102A. Such suggestions are consistent with recent localisations \citep[e.g.][]{Heintz}. 

The X-ray and VHE observations with \textit{Swift} and H.E.S.S. allows us to probe non-thermal persistent emission associated to the FRB host galaxy or its source. Recently, H.E.S.S. observed SGR1935+2154~\citep{hess_sgr1935} that is a Galactic magnetar linked to a repeating FRB and its first X-ray counterpart. Magnetar X-ray flares could in fact be non-thermal in nature~\citep{li2020insighthxmt} indicating the presence of particle acceleration that could  potentially reach the VHE domain. The inverse Compton process is a primary candidate for the production of VHE non-thermal emission. H.E.S.S. observations did not lead to a detection of a persistent or a transient source associated to FRB\,20171019A. We found no X-ray counterparts and thus derived the upper limits to constrain these emissions. In the case of existence of X-ray non-thermal outbursts, the lack of VHE detection could indicate that inverse Compton is weak in the vicinity of the magnetars or that the VHE gamma-ray emission is quenched. This latter scenario could be explained by the fact that inverse Compton is taking place too close to the magnetar's surface, where pair production and photon splitting could be responsible for significant energy losses~\citep{Hu_2019}, preventing energetic particles and photons to reach the nebula.

No persistent emissions were detected towards FRB 20190711A and FRB 20171019A in our MeerKAT observations (see Figures \ref{fig:171019}, \ref{fig:190711}, and \ref{fig:190711_e2}), therefore no follow up observations were conducted towards those FRBs.



\section{Conclusions}\label{sec:conclusions}
Several FRB models envision persistent emission to be associated with these sources. In this paper, we conducted radio observations of three FRBs (FRB 20190714A, 20190711A and 20171019A), and also a multi-wavelength campaign on one of these (FRB 20171019A). 

We detected persistent compact radio emission associated with FRB 20190714A (at $z=0.2365$) using the MeerKAT and e-MERLIN radio telescope. This represents the first detection of the radio continuum emission associated with the host (galaxy) of FRB 20190714A and is only the third known FRB to have such an association. 
We furthermore obtained a radio upper limit of$\sim15\mu$Jy beam$^{-1}$ for the repeating FRBs 20190711A and 20171019A.

We also performed UV, X-ray and VHE observations with the \textit{Swift} and H.E.S.S. instruments and obtained upper limits in the three domains constraining the MWL emissions from FRB\,20171019A. The search for FRB MWL counterparts is ongoing within the H.E.S.S. collaboration and more results will be published in future works.  

Given the association of a comparatively fainter persistent source, FRB 20190714A may potentially be a repeating FRB whose age lies between that of FRB 20121102A and FRB 20180916A.



\section*{Acknowledgements}

This paper makes use of the MeerKAT data (Project ID: SCI-20190418-VC-01). The MeerKAT telescope is operated by the South African Radio Astronomy Observatory, which is a facility of the National Research Foundation, an agency of the Department of Science and Innovation (DSI). This work made use of the Inter-University Institute for Data Intensive Astronomy (IDIA) visualization lab https://vislab.idia.ac.za. IDIA is a partnership of the University of Cape Town, the University of Pretoria, the University of the Western Cape and the South African Radio astronomy Observatory. e-MERLIN is a National Facility operated by the University of Manchester at Jodrell Bank Observatory on behalf of STFC.

The authors acknowledge funding from the European Research Council (ERC) under the European Union’s Horizon 2020 research and innovation programme (grant agreement No 694745). The support of the Namibian authorities and of the University of Namibia in facilitating the construction and operation of H.E.S.S. is gratefully acknowledged, as is the support by the German Ministry for Education and Research (BMBF), the Max Planck Society, the German Research Foundation (DFG), the Helmholtz Association, the Alexander von Humboldt Foundation, the French Ministry of Higher Education, Research and Innovation, the Centre National de la Recherche Scientifique (CNRS/IN2P3 and CNRS/INSU), the Commissariat à l'\'energie atomique et aux \'energies alternatives (CEA), the U.K. Science and Technology Facilities Council (STFC), the Knut and Alice Wallenberg Foundation, the National Science Centre, Poland grant no. 2016/22/M/ST9/00382, the South African Department of Science and Technology and National Research Foundation, the University of Namibia, the National Commission on Research, Science \& Technology of Namibia (NCRST), the Austrian Federal Ministry of Education, Science and Research and the Austrian Science Fund (FWF), the Australian Research Council (ARC), the Japan Society for the Promotion of Science and by the University of Amsterdam. We appreciate the excellent work of the technical support staff in Berlin, Zeuthen, Heidelberg, Palaiseau, Paris, Saclay, T\"ubingen and in Namibia in the construction and operation of the equipment. This work benefited from services provided by the H.E.S.S. Virtual Organisation, supported by the national resource providers of the EGI Federation.

\section*{Data availability}

The data underlying this article will be shared on reasonable request to the corresponding authors.








\appendix

\section{Author affiliations}

$^{1}$Centre for Space Research, North-West University, Potchefstroom 2531, South Africa\\
$^{2}$Department of Physics and Astronomy, Faculty of Physical Sciences, University of Nigeria, Carver Building, 1 University Road, Nsukka 410001, Nigeria\\
$^{3}$Jodrell Bank Centre for Astrophysics, Department of Physics and Astronomy, University of Manchester, Manchester M13 9PL, UK\\
$^{4}$Sydney Institute for Astronomy, School of Physics, The University of Sydney, NSW 2006, Australia \\
$^{5}$Max-Planck-Institut f\"ur Radioastronomie, Auf dem H\"ugel 69, D-53121 Bonn, Germany\\
$^{6}$IRFU, CEA, Universit\'e Paris-Saclay, F-91191 Gif-sur-Yvette, France\\
$^{7}$Department of Physics and Electronics, Rhodes University, PO Box 94, Grahamstown 6140, South Africa\\
$^{8}$South African Radio Astronomy Observatory, Black River Park, 2 Fir Street, Observatory, Cape Town 7925, South Africa\\
$^{9}$Astrophysics, Department of Physics, University of Oxford, Keble Road, Oxford OX1 3RH, UK \\ 
$^{10}$National University of Ireland Galway, University Road, Galway, H91 TK33, Ireland \\
$^{11}$SKA Observatory, Jodrell Bank Observatory, Macclesfield, Cheshire SK11 9DL, UK \\
$^{12}$Dublin Institute for Advanced Studies, 31 Fitzwilliam Place, Dublin 2, Ireland \\
$^{13}$Max-Planck-Institut f\"ur Kernphysik, P.O. Box 103980, D 69029 Heidelberg, Germany \\
$^{14}$High Energy Astrophysics Laboratory, RAU,  123 Hovsep Emin St  Yerevan 0051, Armenia \\
$^{15}$Landessternwarte, Universit\"at Heidelberg, K\"onigstuhl, D 69117 Heidelberg, Germany \\
$^{16}$Aix Marseille Universit\'e, CNRS/IN2P3, CPPM, Marseille, France \\
$^{17}$Laboratoire Leprince-Ringuet, École Polytechnique, CNRS, Institut Polytechnique de Paris, F-91128 Palaiseau, France \\
$^{18}$University of Namibia, Department of Physics, Private Bag 13301, Windhoek 10005, Namibia \\
$^{19}$Instytut Fizyki J\c{a}drowej PAN, ul. Radzikowskiego 152, 31-342 Krak{\'o}w, Poland \\
$^{20}$DESY, D-15738 Zeuthen, Germany \\
$^{21}$School of Physics, University of the Witwatersrand, 1 Jan Smuts Avenue, Braamfontein, Johannesburg, 2050 South Africa \\
$^{22}$Université de Paris, CNRS, Astroparticule et Cosmologie, F-75013 Paris, France \\
$^{23}$Department of Physics and Electrical Engineering, Linnaeus University,  351 95 V\"axj\"o, Sweden \\
$^{24}$Laboratoire Univers et Théories, Observatoire de Paris, Université PSL, CNRS, Université de Paris, 92190 Meudon, France \\
$^{25}$Sorbonne Universit\'e, Universit\'e Paris Diderot, Sorbonne Paris Cit\'e, CNRS/IN2P3, Laboratoire de Physique Nucl\'eaire et de Hautes Energies, \\LPNHE, 4 Place Jussieu, F-75252 Paris, France \\
$^{26}$Université Savoie Mont Blanc, CNRS, Laboratoire d'Annecy de Physique des Particules - IN2P3, 74000 Annecy, France \\
$^{27}$Astronomical Observatory, The University of Warsaw, Al. Ujazdowskie 4, 00-478 Warsaw, Poland \\
$^{28}$Friedrich-Alexander-Universit\"at Erlangen-N\"urnberg, Erlangen Centre for Astroparticle Physics, Erwin-Rommel-Str. 1, D 91058 Erlangen, Germany \\
$^{29}$University of Oxford, Department of Physics, Denys Wilkinson Building, Keble Road, Oxford OX1 3RH, UK \\
$^{30}$Universit\'e Bordeaux, CNRS/IN2P3, Centre d'\'Etudes Nucl\'eaires de Bordeaux Gradignan, 33175 Gradignan, France \\
$^{31}$Institut f\"ur Physik und Astronomie, Universit\"at Potsdam,  Karl-Liebknecht-Strasse 24/25, D 14476 Potsdam, Germany \\
$^{32}$Obserwatorium Astronomiczne, Uniwersytet Jagiello{\'n}ski, ul. Orla 171, 30-244 Krak{\'o}w, Poland \\
$^{33}$Institute of Astronomy, Faculty of Physics, Astronomy and Informatics, Nicolaus Copernicus University,  Grudziadzka 5, 87-100 Torun, Poland \\
$^{34}$Nicolaus Copernicus Astronomical Center, Polish Academy of Sciences, ul. Bartycka 18, 00-716 Warsaw, Poland \\
$^{35}$Institut f\"ur Astronomie und Astrophysik, Universit\"at T\"ubingen, Sand 1, D 72076 T\"ubingen, Germany \\
$^{36}$Institut f\"ur Physik, Humboldt-Universit\"at zu Berlin, Newtonstr. 15, D 12489 Berlin, Germany \\
$^{37}$Laboratoire Univers et Particules de Montpellier, Universit\'e Montpellier, CNRS/IN2P3,  CC 72, Place Eug\`ene Bataillon, F-34095 Montpellier Cedex 5, France \\
$^{38}$Institut f\"ur Astro- und Teilchenphysik, Leopold-Franzens-Universit\"at Innsbruck, A-6020 Innsbruck, Austria \\
$^{39}$Department of Physics and Astronomy, The University of Leicester, University Road, Leicester, LE1 7RH, United Kingdom \\
$^{40}$GRAPPA, Anton Pannekoek Institute for Astronomy, University of Amsterdam,  Science Park 904, 1098 XH Amsterdam, The Netherlands \\
$^{41}$School of Physical Sciences, University of Adelaide, Adelaide 5005, Australia \\
$^{42}$Yerevan Physics Institute, 2 Alikhanian Brothers St., 375036 Yerevan, Armenia \\
$^{43}$Kavli Institute for the Physics and Mathematics of the Universe (WPI), The University of Tokyo Institutes for Advanced Study (UTIAS), \\The University of Tokyo, 5-1-5 Kashiwa-no-Ha, Kashiwa, Chiba, 277-8583, Japan \\
$^{44}$Department of Physics, Konan University, 8-9-1 Okamoto, Higashinada, Kobe, Hyogo 658-8501, Japan \\
$^{45}$RIKEN, 2-1 Hirosawa, Wako, Saitama 351-0198, Japan



\bsp	
\label{lastpage}
\end{document}